\documentclass[ALICE,manyauthors]{cernphprep}

 \usepackage{graphics}
 \usepackage{graphicx}
\usepackage{color}

\usepackage{amssymb}
\usepackage{multicol}

\usepackage{subfigure} 

\usepackage{lineno}
\usepackage[english]{babel}
\usepackage{hyperref}
 


\begin{document}

\begin{titlepage}

\PHnumber{2010-066}                 
\PHdate{2 December 2010}              


~

~

~

\title{Suppression of charged particle production at large transverse momentum 
in central Pb--Pb collisions at $\sqrt{s_{_{NN}}} = 2.76$~TeV
}
\ShortTitle{Suppression of charged particle production}
\Collaboration{ALICE Collaboration%
         \thanks{See Appendix~\ref{app:collab} for the list of collaboration 
                      members}}
\ShortAuthor{ALICE Collaboration}      







\begin{abstract}
Inclusive transverse momentum spectra of primary charged particles
in Pb--Pb collisions at 
$\sqrt{s_{_{NN}}}$ = 2.76~TeV have been measured
by the ALICE Collaboration at the LHC. 
The data are presented for central and peripheral collisions,
corresponding to 0--5\% and 70--80\% of the hadronic Pb--Pb cross
section.
The measured charged particle
spectra in $|\eta|<0.8$ and $0.3 < p_T <  20$ GeV/$c$
are compared to the expectation in pp collisions at the 
same $\sqrt{s_{_{NN}}}$,
scaled by the number of 
underlying nucleon--nucleon collisions.
The comparison is expressed in terms of the nuclear modification factor 
$R_{AA}$. The result indicates 
only weak medium effects ($R_{AA} \approx $~0.7) in peripheral collisions.
In central collisions, $R_{AA}$ reaches a minimum of about 0.14 at 
$p_T=6$--$7$~GeV/$c$ and increases significantly at larger $p_T$. 
The measured suppression of high--$p_T$ particles is stronger 
than that observed at lower collision energies, indicating that a 
very dense 
medium is formed in central Pb--Pb collisions at the LHC.
\end{abstract}



\end{titlepage}

\setcounter{page}{2}




High energy heavy-ion collisions enable the study of
strongly interacting matter
under extreme conditions. At sufficiently high collision energies 
Quantum-Chromodynamics (QCD) predicts that hot and dense deconfined matter,
commonly referred to as the Quark-Gluon Plasma (QGP), is formed.
With the advent of a new generation of experiments at the 
CERN Large Hadron Collider (LHC)~\cite{LHC} a new 
energy domain is accessible to study the properties of this state.

Previous experiments at the Relativistic Heavy Ion Collider (RHIC) reported 
that hadron production at high transverse momentum ($p_T$) 
in central (head-on) Au--Au collisions at a centre-of-mass 
energy  per
nucleon pair $\sqrt{s_{_{NN}}}$ of 200~GeV
is suppressed by a factor 4--5 compared to 
expectations from an independent 
superposition of nucleon-nucleon (NN)
collisions~\cite{WPBrahms,WPPhobos,WPStar,WPPhenix}. 
The dominant production mechanism for high-$p_T$ hadrons is the fragmentation 
of high-$p_T$ partons that originate in hard scatterings in the 
early stage of the nuclear collision. 
The observed suppression at RHIC is generally attributed to energy 
loss of the partons as they propagate through the hot and dense QCD 
medium~\cite{wang,Baier,wicks,vitev,eskola}.

To quantify nuclear medium effects at high $p_T$, the so called
{\it nuclear modification factor} $R_{AA}$ 
is used. $R_{AA}$ is defined as the ratio 
of the charged particle yield in Pb--Pb to that in pp, 
scaled by the number of binary nucleon--nucleon collisions 
$\langle N_{\rm coll}\rangle$
\begin{displaymath} 
\nonumber
R_{AA}(p_T)\,=\,\frac{(1/N_{evt}^{AA})\,d^2N^{AA}_{\rm ch}/d\eta dp_T}
{\langle N_{\rm coll} \rangle\,(1/N_{evt}^{pp})\,d^2N^{pp}_{\rm ch}/d\eta dp_T},
\end{displaymath}
where $\eta=-\ln (\tan \theta/2)$ is the pseudo-rapidity and
$\theta$ is the polar angle 
between the charged particle direction and the beam axis.
The number of binary nucleon--nucleon collisions 
$\langle N_{\rm coll}\rangle$ is given by the product
of the nuclear overlap function 
$\langle T_{AA} \rangle$~\cite{glauber} and
the inelastic NN cross section $\sigma^{NN}_{\rm inel}$.
If no nuclear modification is present, $R_{AA}$ is unity at
high $p_T$.

At the larger LHC energy the density of the medium is expected to be
higher than at RHIC, leading to a larger energy loss of high 
$p_T$ partons.
On the other hand, the less steeply falling spectrum at the higher energy
will lead to a smaller suppression in the $p_T$ spectrum of 
charged particles, 
for a given magnitude of partonic energy loss~\cite{vitev,eskola}. 
Both the value of $R_{AA}$ in central collisions as well as its 
$p_T$ dependence may also in part be influenced by gluon shadowing and 
saturation effects, which in general decrease with increasing $x$ and $Q^2$.


This Letter reports the measurement of the inclusive 
primary charged particle transverse momentum distributions at 
mid-rapidity in central and peripheral
Pb--Pb collisions at $\sqrt{s_{_{NN}}}=2.76$~TeV 
by the ALICE experiment~\cite{ALICE-det}. 
Primary particles are defined as prompt
particles produced in the collision, including decay products, 
except those from 
weak decays of strange particles.
The data were collected in the first heavy-ion collision 
period at the LHC. 
A detailed description of the  
experiment can be found in~\cite{ALICE-det}.

For the present analysis, charged particle tracking utilizes
the Inner Tracking System (ITS) and the
Time Projection Chamber (TPC)~\cite{TPC-nim}, both of which cover 
the central region in the pseudo-rapidity range $|\eta|<0.9$.
The ITS and TPC detectors are located in the ALICE central 
barrel and operate in the 0.5~T magnetic
field of a large solenoidal magnet.
The TPC is a cylindrical drift detector with
two readout planes on the endcaps.
The active volume covers $85<r<247$~cm and 
$-250<z<250$~cm in the radial and longitudinal directions, respectively. 
A high voltage membrane at $z=0$ divides the active volume into two halves
and provides the electric drift field of 400~V/cm, resulting in a maximum
drift time of 94~$\mu$s.

The ITS is used for charged particle tracking and trigger purposes.
It is composed of six cylindrical layers of high resolution
silicon tracking detectors with radial distances 
to the beam line from 3.9 to 43~cm. 
The two innermost layers are the Silicon Pixel Detectors (SPD) 
with a total of 9.8~million pixels, read out by 1200 chips.
Each chip provides a fast signal if at least one
of its pixels is hit. The signals from the 1200 chips are
combined in a programmable logic unit which supplies a
trigger signal. The SPD contributes to the minimum-bias trigger,
if hits are detected on at least two chips on the outer layer.
The SPD is followed by two layers of Silicon Drift Detectors (SDD) 
with 133k readout channels. The two outermost layers are Silicon
Strip Detectors (SSD) consisting of double-sided silicon micro-strip
sensors, for a total of 2.6~million readout channels.

The two forward scintillator hodoscopes (VZERO-A and VZERO-C)
cover the pseudo-rapidity ranges $2.8<\eta<5.1$ and 
$-3.7<\eta<-1.7$. The sum of the amplitudes of the signals in the 
VZERO scintillators is used as a measure for the event centrality.
The VZERO detectors also provide a fast trigger signal if at least
one particle hit was detected.

During the heavy-ion data-taking period,
up to 114 bunches, each containing about
$7\times10^{7}$ ions of $^{208}$Pb, were collided at 
$\sqrt{s_{_{NN}}}=2.76$~TeV in the ALICE interaction region.
The rate of hadronic events was about 100~Hz, 
corresponding to an estimated luminosity of 
1.3$\times$10$^{25}$~cm$^{-2}$s$^{-1}$. 
The detector readout was triggered by the LHC bunch-crossing signal
and a minimum-bias
interaction trigger based on
trigger signals from VZERO-A, VZERO-C, and SPD. 
The present analysis combines runs taken with two different
minimum-bias conditions. In the first set of runs, two out of the 
three trigger signals were required, while in the second set a 
coincidence
between VZERO-A and VZERO-C was used. Both trigger conditions
have similar efficiency
for hadronic interactions, but the latter suppresses a large 
fraction of electromagnetic reactions.

\begin{figure}
\centering
\includegraphics[width=7.8cm]{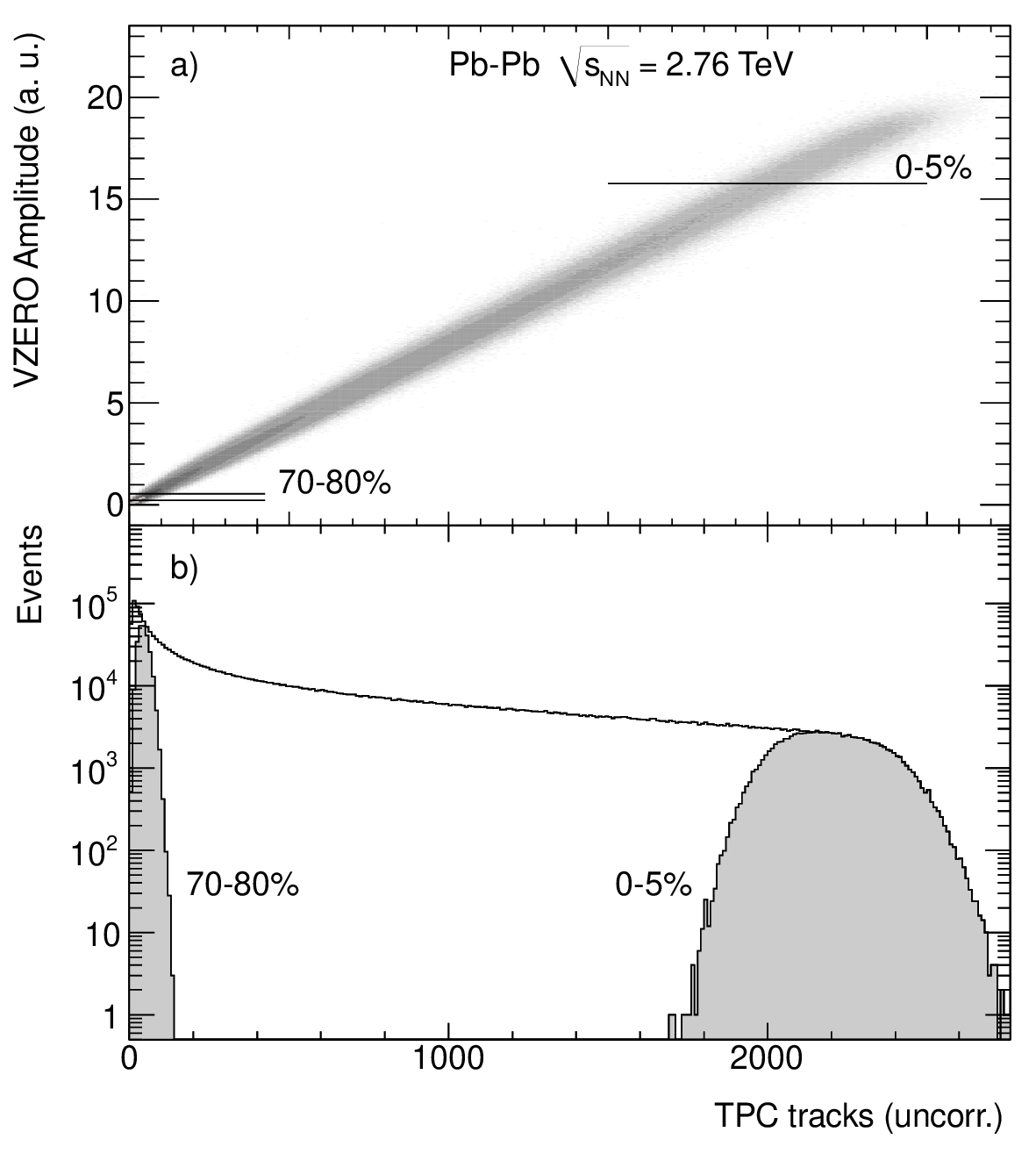}
\caption{\label{fig1} Upper panel: Correlation between VZERO amplitude and
the uncorrected track multiplicity in the TPC. Indicated are the cuts
for the centrality ranges used in this analysis. 
Lower panel: Minimum-bias
distribution of the TPC track multiplicity. The central 
(0--5\%) and peripheral
(70--80\%) event subsamples used for this analysis are shown as grey 
histograms.}
\end{figure}

The following analysis is based on $2.3\times 10^6$ minimum-bias 
Pb--Pb events, which passed the offline
event selection. This selection is based on VZERO
timing information and the correlation between TPC tracks and
hits in the SPD to reject background events
coming from parasitic beam interactions.
Additionally, a minimal energy deposit in the Zero Degree Calorimeters
(ZDC) is required to
further suppress electromagnetic interactions. Only
events with reconstructed vertex at $|z_{\rm vtx}|<10$~cm 
were used.
The definition of the event centrality is based on the sum of
the amplitudes measured in the VZERO detectors as described 
in~\cite{paper-mult1}. 
Alternative centrality measures utilize the cluster multiplicity 
in the outer layer of the SPD or the multiplicity of 
reconstructed tracks.
The correlation between the VZERO amplitude and the uncorrected TPC track 
multiplicity in $|\eta|<0.8$ is illustrated in Fig.\ref{fig1}. 
The VZERO amplitude distribution is fitted using a Glauber 
model~\cite{glauber}
to determine percentage intervals of the hadronic cross section, 
as described in~\cite{paper-mult1}.
We used a Glauber model Monte Carlo simulation assuming 
$\sigma^{NN}_{\rm inel}= 64$~mb,
a Woods-Saxon nuclear density with radius 
$6.62\pm0.06$~fm and surface
diffuseness $0.546\pm0.010$~fm~\cite{WS-parameters}. 
A minimum inter-nucleon distance of
$0.4\pm0.4$~fm is assumed.
The Glauber Monte Carlo allows one to relate the event 
classes to the mean numbers
of participating nucleons $\langle N_{\rm part}\rangle$ and 
binary collisions 
$\langle N_{\rm coll}\rangle$ (see Table~\ref{tab1})
by geometrically ordering events according to the impact parameter 
distribution. 
The errors include the experimental
uncertainties in the parameters used in the
Glauber simulation and an uncertainty of $\pm5$~mb in $\sigma^{NN}_{\rm inel}$.
The TPC multiplicity distributions for the central and peripheral 
event samples selected for this analysis, corresponding 
to the 0--5\% and 70--80\% most central 
fraction of the hadronic Pb--Pb cross 
section, are shown in the lower panel of Fig.~\ref{fig1}.
\begin{table}
\centering
\caption{\label{tab1} 
The average numbers of participating
nucleons $\langle N_{\rm part}\rangle$, binary nucleon--nucleon 
collisions $\langle N_{\rm coll}\rangle$, and the average nuclear 
overlap function $\langle T_{AA}\rangle$ for the two centrality
bins, expressed in percentages of the hadronic cross section.}
~\\
~\\
\begin{tabular}{cccc}
\hline
\hline
Centrality&$\langle N_{\rm part}\rangle$& 
$\langle N_{\rm coll}\rangle$&$\langle T_{AA}\rangle$(mb$^{-1})$ \\
\hline
0--5\% & $383\pm2$ & $1690\pm131$ & $26.4\pm0.5$\\
70--80\% & $15.4\pm0.4$ & $15.7\pm0.7$ & $0.25\pm0.01$\\
\hline
\hline
\end{tabular}
\end{table} 
Charged particle tracks are reconstructed in the ITS and TPC 
detectors. 
Track candidates in the TPC are selected in the pseudo-rapidity range 
$|\eta|<0.8$. 
Track quality cuts in the TPC are based on the number of
reconstructed space points (at least 70 out of a maximum of 159) and 
the $\chi^2$ per space point of the momentum fit (lower than 4). 
The TPC track candidates are projected to the ITS and 
used for further analysis,
if at least two matching hits in the ITS are found, 
including at least one in the SPD.
The average number of associated hits in the ITS is 4.7 
for the selected tracks.
The event vertex is reconstructed by extrapolating the particle
tracks to the interaction region. 
The event vertex reconstruction is fully efficient
in both the peripheral and the central event sample.
Tracks are rejected from the final sample if their distance of closest
approach to the reconstructed vertex 
in longitudinal and radial direction, $d_{z}$ and $d_{xy}$, satisfies 
$d_{z}>2\,{\rm cm}$ or 
$d_{xy} > 0.018\,{\rm cm}+0.035\,{\rm cm}\cdot p_T^{-1.01} $,
with $p_T$ in GeV/$c$.

The efficiency and purity of primary charged particles using these
cuts are estimated using a Monte Carlo simulation including 
HIJING~\cite{hijing} 
events 
and a GEANT3~\cite{geant} model of the detector response~\cite{aliroot}. 
We used a HIJING tune which reproduces approximately the measured
charged particle density in central collisions~\cite{paper-mult1}.
In central events, the overall primary charged particle 
efficiency in $|\eta|<0.8$ is 60\% at $p_T=0.3$~GeV/$c$
and increases to 65\% at $p_T =0.6$~GeV/$c$ and above.
In peripheral events, the efficiency is larger by about 2--3\%.
The contamination from secondaries is 6\% at $p_T=0.3$~GeV/$c$
and decreases to about 2\% at $p_T>1$~GeV/$c$, with no significant
centrality dependence.
This contribution was estimated using the $d_{xy}$ distributions of 
data and HIJING and is consistent 
with a first estimate
of the strangeness to charged particle ratio from the reconstruction 
of K$^0_s$, $\Lambda$ and $\bar{\Lambda}$ invariant mass peaks. 

The momentum of charged particles is reconstructed from the track
curvature measured in the ITS and TPC. The momentum resolution 
can be parametrized as
$(\sigma(p_T)/p_T)^2=a^2+(b\cdot p_T)^2$. It is estimated
from the track residuals to the momentum fit and verified by cosmic 
muon events 
and the width of the
invariant mass peaks of $\Lambda$, $\bar{\Lambda}$ and K$^0_s$.
While $a=0.01$ for both centrality bins, there is a weak centrality 
dependence of $b$, i.e. $b=0.0045$~(GeV/$c$)$^{-1}$ in peripheral events and 
$b=0.0056$~(GeV/$c$)$^{-1}$ in central events. This is related 
to a slight decrease for more central events
of the average number of space points
in the TPC.
The modification of the spectra arising from the finite momentum 
resolution is estimated by
Monte Carlo. 
It results in an  overestimate of the
yield by up to 8\% at $p_T=20$~GeV/$c$ in central events. 
This was accounted for by introducing a $p_T$ dependent 
correction factor to the $p_T$ spectra.
From the mass difference between $\Lambda$ and $\bar{\Lambda}$ and the
ratio of positive over negative charged tracks, assuming charge
symmetry at high $p_T$, the
upper limit of the systematic uncertainty of the momentum scale 
is estimated to be $|\Delta(p_T)/p_T|<0.002$. This has negligible
effect on the measured spectra.

\begin{table}
\centering
\caption{\label{tab2} Contributions to the systematic 
uncertainties on the inclusive
spectra. For the $p_T$ dependent errors the ranges are given.}
~\\
~\\
\begin{tabular}{lcc}
\hline
\hline
Centrality class&  0--5\% & 70--80\%  \\
\hline
Centrality selection & 1\% & 7\% \\
Track and event selection cuts & 1--4\% & 1--4\%\\
Particle composition & 1--4\% & 1--4\% \\
Material budget & 1--2\% & 1--2\%\\
Secondary particle rejection & $<$1\% & $<$1\% \\
Tracking efficiency & 2--6\% & 2--6\% \\
\hline
Total systematic uncertainties & 5--7\% & 8--10\% \\
\hline
\hline
\end{tabular}
\end{table} 

Table~\ref{tab2} shows the systematic uncertainties 
obtained by a comparison 
of different centrality measures (using the SPD instead of VZERO), 
and by varying the track and event
quality cuts and the Monte Carlo assumptions.
In particular, we studied a variation of the most abundant charged
particle species (p, $\pi$, K) by $\pm30$\%, the material budget
by $\pm$7\%, and the secondary yield from strangeness decays in the
Monte Carlo by $\pm 30$\%. 
We have used the differences between the standard analysis and one 
based only on the use of TPC tracks to estimate the uncertainty
on the track efficiency corrections, included in the systematic errors.
The total systematic uncertainties on the corrected $p_T$
spectra depend on $p_T$ and are 8--10\% and 5--7\% for 
the peripheral and central event samples, respectively.

The determination of $R_{AA}$ requires a 
pp reference at $\sqrt{s}=2.76$~TeV, where
no pp measurement exists.
Different approaches are at hand which allow a prediction of
the $p_T$ spectrum at a given $\sqrt{s}$ by scaling
existing data at different energies. Such approaches assume
general scaling properties of perturbative QCD (pQCD) or rely on 
next-to-leading order (NLO) pQCD
calculations. 
The present analysis follows a data-driven approach with minimal 
theoretical assumptions where,
in order to minimize systematic uncertainties,
only measurements by ALICE are considered.
In this approach, the pp reference spectrum is obtained
by interpolating the differential yields 
$d^2 N_{ch}^{pp}/d\eta d p_T$ of charged particles measured 
in inelastic pp collisions at $\sqrt{s}=$~0.9 and 7~TeV by 
ALICE~\cite{paper3,7TeV}.
The interpolation is performed in bins of $p_T$, based on the assumption 
that the increase of the yield with $\sqrt{s}$ follows a power law.
Above $p_T=2$~GeV/$c$, the measured spectra at the two energies 
are parametrized by a modified Hagedorn 
function~\cite{hagedorn} and a power law 
to reduce bin-by-bin fluctuations.
Systematic uncertainties on the pp reference spectrum
arise from the experimental
errors of the measured spectra at 0.9 and 7 TeV, from the
parametrization, and from the interpolation procedure in $\sqrt{s}$.
The combined statistical and systematic data errors result in
a 9--10\% uncertainty on the pp reference spectrum at $\sqrt{s}=2.76$~TeV, 
depending on $p_T$. 
The interpolation procedure was verified using PHOJET~\cite{phojet} 
and PYTHIA~\cite{pythia}
(tunes D6T~\cite{d6t} and Perugia0~\cite{perugia0}) at 0.9, 2.76 
and 7~TeV. The generated and 
interpolated spectra 
at 2.76~TeV agree within the quoted uncertainties.
Finally, the scaled pp yield in a given centrality class is obtained by 
multiplication of the pp reference spectrum with 
$\langle N_{\rm coll}\rangle$, see Table~\ref{tab1}.
The uncertainty in 
$\langle N_{\rm coll} \rangle$ 
results in an additional $p_T$-independent scaling uncertainty on the 
scaled pp reference. 

Alternative approaches to derive the pp reference spectrum 
are investigated to study the sensitivity of $R_{AA}$
to the specific choice of our method.
Replacing in the interpolation the $p_T$ spectrum at 0.9~TeV by the one 
measured in p$\bar{\rm p}$ 
at $\sqrt{s}=1.96$~TeV in $|\eta|<1$ by the CDF collaboration~\cite{cdf} 
results in a pp reference
spectrum which is 5--15\% lower than the reference spectrum
described above.
A different procedure to obtain a pp reference 
is based on a scaling of the $p_T$ spectra at 0.9 or 7 TeV
to 2.76 TeV by the relative $\sqrt{s}$ dependence predicted 
by NLO pQCD calculations~\cite{stratmann} (referred to as ``NLO scaling'').
Using the 7 TeV spectrum as a starting point, good agreement 
with the reference obtained from interpolation is found.
Starting instead from 0.9 TeV results in a spectrum which is 30--50\%
higher than the interpolation reference.
The pp reference spectra derived from the use of the CDF data 
in the interpolation and from NLO scaling of the 0.9 TeV data 
are used in the following to illustrate 
the dependence of $R_{AA}$ at high $p_T$ 
on the choice of the reference spectrum.
\begin{figure}
\centering
\includegraphics[width=7.8cm]{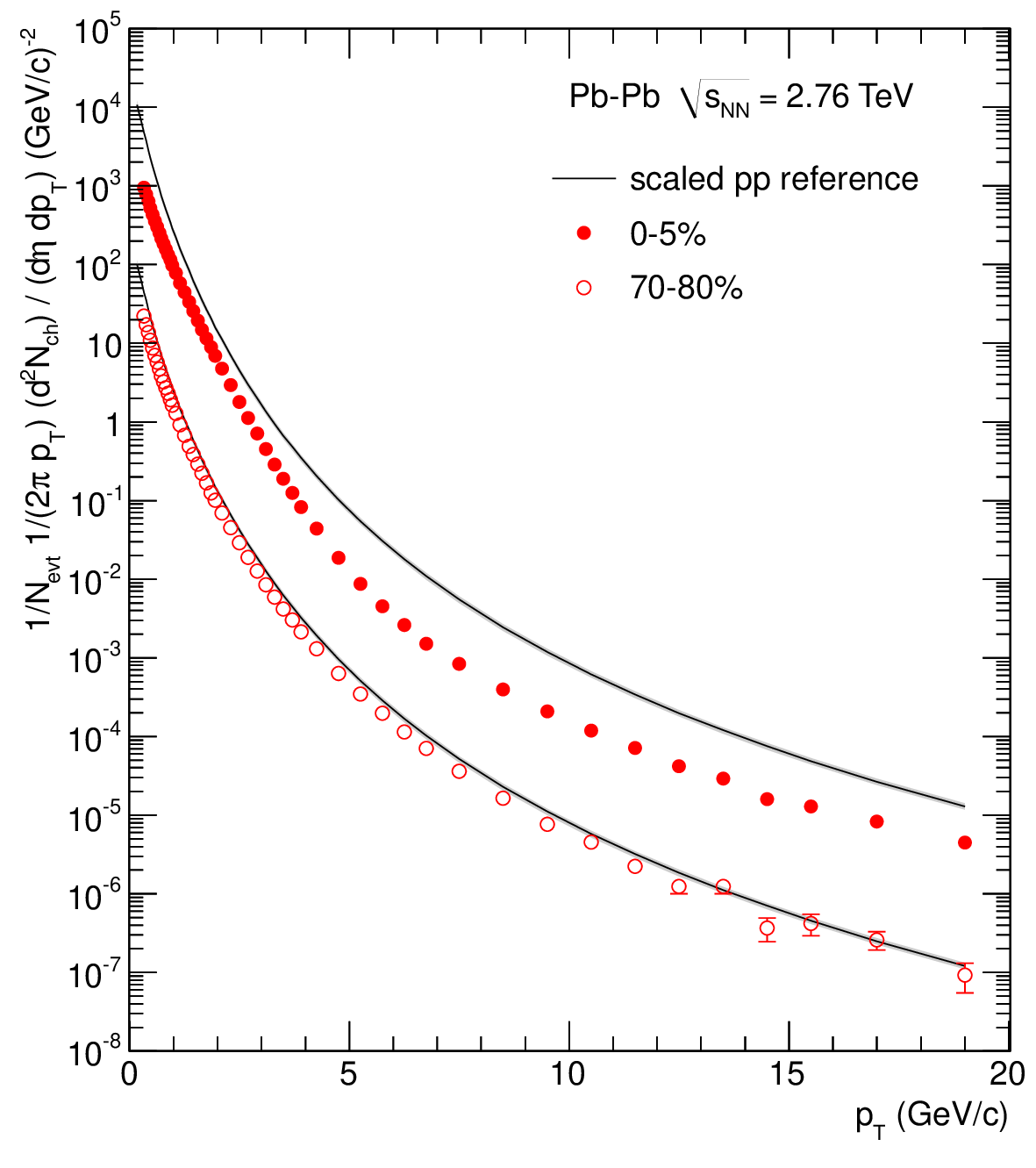}
\caption{\label{fig2} The $p_T$ distributions of primary charged particles
at mid-rapidity ($|\eta|<0.8$) in central (0--5\%) and peripheral 
(70--80\%)
Pb--Pb collisions at $\sqrt{s_{_{NN}}}=2.76$~TeV. Error bars are statistical
only. The systematic data errors are smaller than the symbols.
The scaled pp references are shown as the two curves, the upper
for 0--5\% centrality and the lower for 70--80\%. 
The systematic uncertainties of the pp reference spectra are contained 
within the thickness of the line.}
\end{figure}

The $p_T$ distributions of primary charged particles in central and peripheral
 Pb--Pb collisions
at 2.76~TeV are shown in Fig.~\ref{fig2}, together with the binary-scaled
yields from pp collisions. 
The $p_{T}$-dependence is similar for
the pp reference and for
peripheral Pb--Pb collisions, exhibiting a power law behaviour at 
$p_T>3$~GeV/$c$, which is characteristic of perturbative 
parton scattering
and vacuum fragmentation. 
In contrast, the spectral shape in central collisions clearly deviates 
from the scaled pp reference and is closer to an exponential 
in the $p_T$ range below 5~GeV/$c$. 

\begin{figure}
\centering
\includegraphics[width=7.8cm]{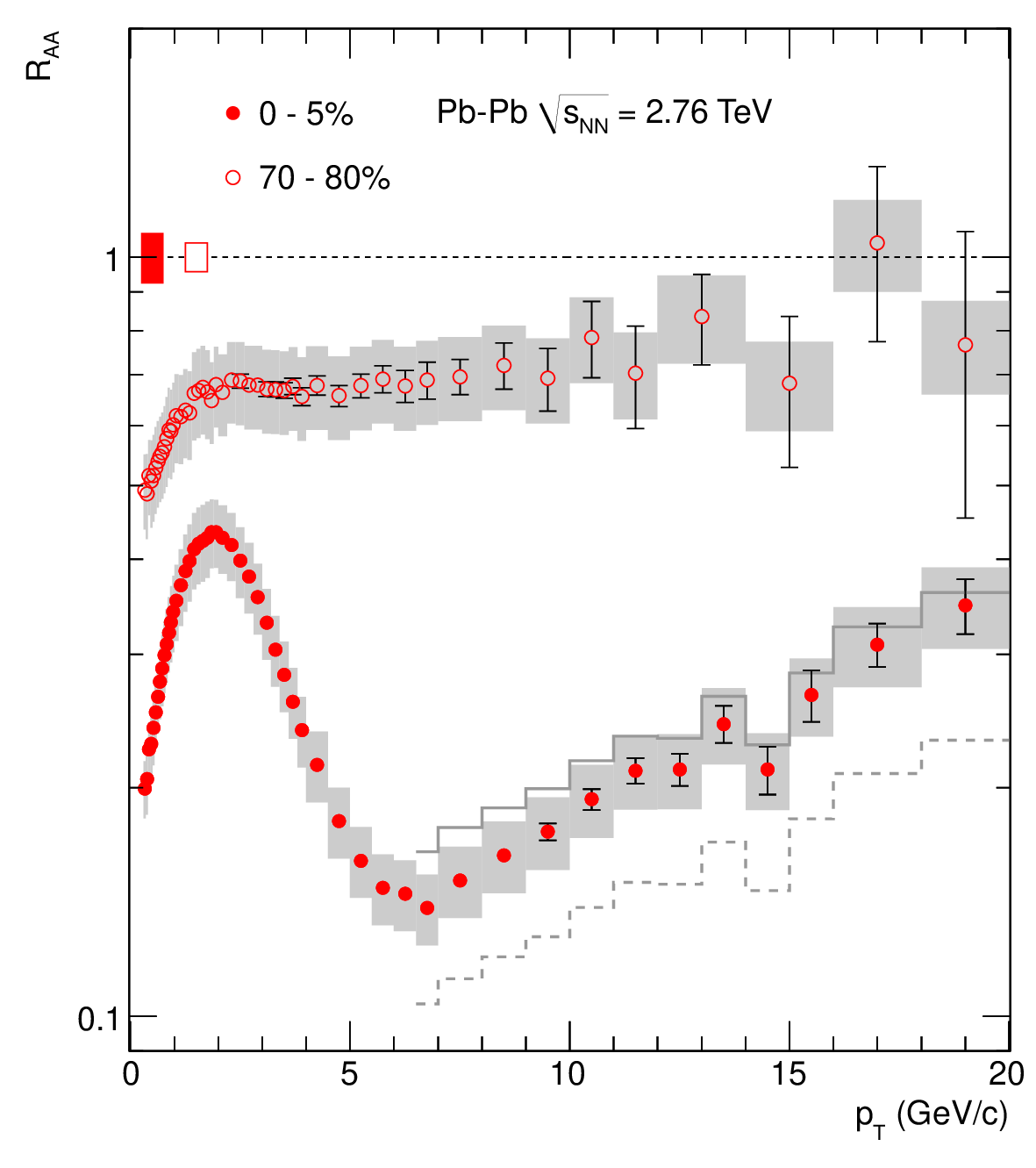}
\caption{\label{fig3} $R_{AA}$ in central (0--5\%) and 
peripheral (70--80\%)
Pb--Pb collisions at $\sqrt{s_{_{NN}}}=2.76$~TeV. Error bars indicate
the statistical uncertainties. The boxes contain the systematic errors
in the data
and the $p_T$ dependent systematic errors on the pp reference, added
in quadrature. The histograms indicate, for central collisions only,
the result for
$R_{AA}$ at $p_T>6.5$~GeV/$c$ 
using alternative pp references obtained by the 
use of the p$\bar{\rm p}$ 
measurement at $\sqrt{s_{_{NN}}}=1.96$~TeV~\cite{cdf}
in the interpolation procedure (solid) and by applying NLO scaling 
to the pp data at 0.9 TeV (dashed) (see text).
The vertical bars around $R_{AA}=1$ show the $p_T$ independent 
uncertainty on $\langle N_{\rm{ coll}} \rangle$.
}
\end{figure}

Figure~\ref{fig3} shows the nuclear modification factor 
$R_{AA}$ for central and peripheral Pb--Pb collisions. 
The nuclear modification factor deviates from one in 
both samples. 
At high $p_T$, where production from hard processes is expected to 
dominate, there is a marked difference between peripheral and central 
events. In peripheral collisions, the nuclear modification factor 
reaches about 0.7 and shows no pronounced $p_T$ dependence 
for $p_T>2$~GeV/$c$. 
In central collisions, $R_{AA}$ is again significantly different from one, 
reaching a minimum of $R_{AA}\approx 0.14$ at $p_T = 6$--$7$ GeV/$c$. 
In the intermediate region there is a strong dependence on $p_T$ with 
a maximum at $p_T = 2$~GeV/$c$. This may reflect a variation of the 
particle composition in heavy-ion collisions with respect to pp, 
as observed at RHIC~\cite{Abelev:2006jr,Adler:2003cb}. 
A significant rise of $R_{AA}$ by about a factor of two is observed
for $7<p_T<20$~GeV/$c$.
Shown as histograms in Fig.~\ref{fig3}, for central events only, are 
the results for $R_{AA}$ at high $p_T$,
using alternative
procedures for the computation of the pp reference, as described
above. 
For such scenarios, the overall value for $R_{AA}$ is shifted,
but a significant increase of $R_{AA}$ in central collisions
for $p_T > 7$~GeV/$c$ persists.

\begin{figure}
\centering
\includegraphics[width=7.8cm]{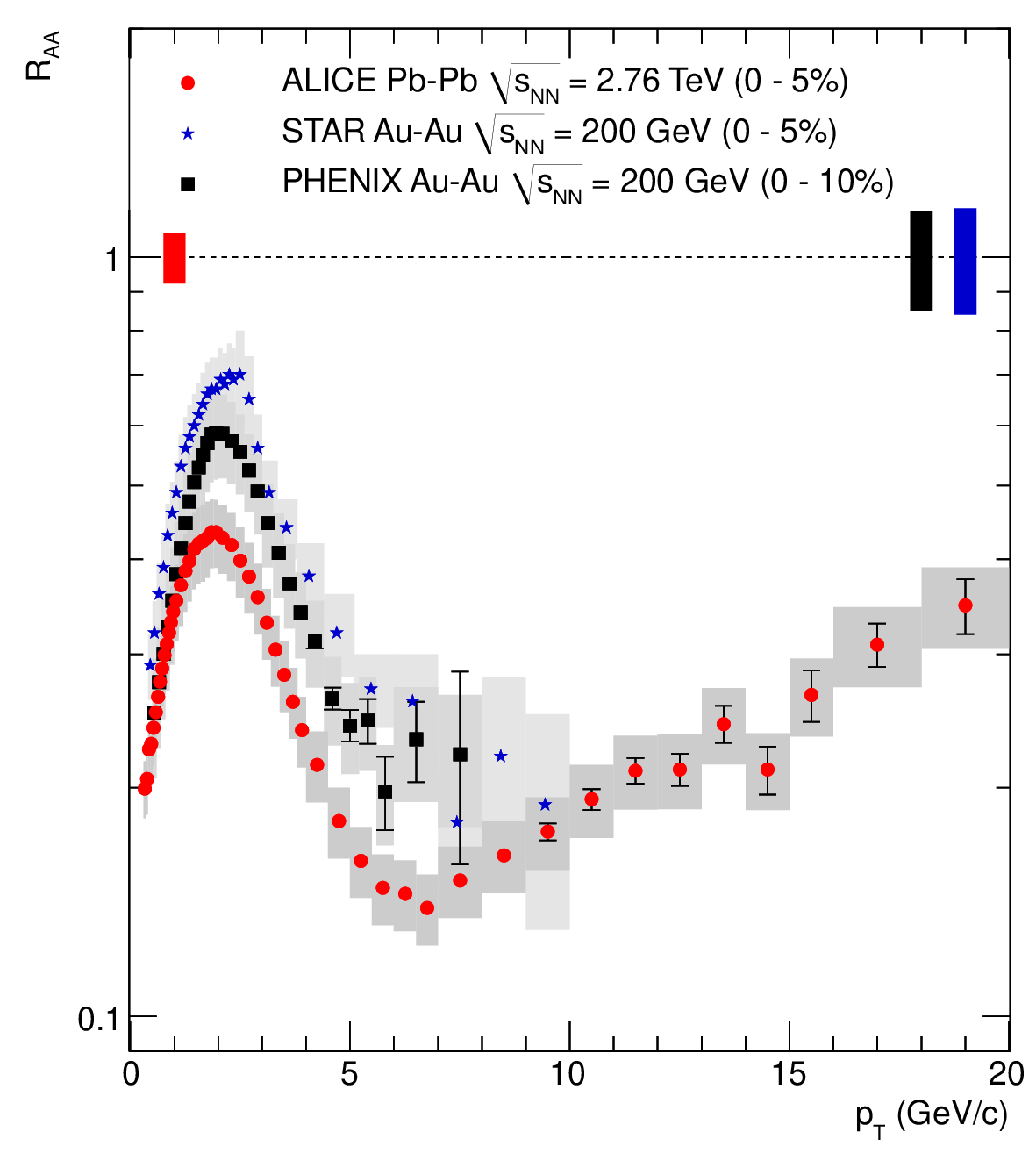}
\caption{\label{fig4} Comparison of $R_{AA}$ in central Pb--Pb
collisions at LHC to measurements at $\sqrt{s_{_{ NN}}}=200$~GeV
by the PHENIX~\cite{Adler:2003au} and 
STAR~\cite{Adams:2003kv} experiments at RHIC. The error representation 
of the ALICE data is as in Fig.~\ref{fig3}. The statistical and systematic
errors of the PHENIX data are shown as error bars and boxes, respectively.
The statistical and systematic errors of the STAR data are combined and 
shown as boxes. The vertical bars around $R_{AA}=1$ indicate the $p_T$ 
independent scaling errors on $R_{AA}$.}
\end{figure}

In Fig.~\ref{fig4} the ALICE result in central Pb--Pb collisions 
at the LHC
is compared to measurements of $R_{AA}$ of charged hadrons 
($\sqrt{s_{_{NN}}} = 200$ GeV) by the PHENIX and STAR 
experiments~\cite{Adler:2003au,Adams:2003kv} at RHIC. 
At 1~GeV/$c$ the measured value of $R_{AA}$ 
is similar to those from RHIC. 
The position and shape of the maximum at $p_T
\sim 2$ GeV/$c$ and the subsequent decrease are similar at RHIC
and LHC, contrary to expectations from a recombination 
model~\cite{Fries:2003fr}.
Despite the much flatter $p_T$ spectrum in pp at the LHC,
the nuclear modification factor at $p_T=6$--$7$~GeV/$c$ 
is smaller than 
at RHIC. 
This suggests an enhanced energy loss at LHC and therefore a denser medium.
A quantitative determination of the energy loss and medium density
will require further investigation of gluon shadowing and saturation
in the present energy range and detailed theoretical modeling.

In summary, we have measured the primary charged particle $p_T$ 
spectra and nuclear
modification factors $R_{AA}$ in central (0--5\%) and peripheral
(70--80\%) Pb--Pb collisions at $\sqrt{s_{_{NN}}}=2.76$ TeV
with the ALICE experiment. The nuclear modification factor in peripheral
collisions is large and
independent of $p_T$ for $p_{T}>2$ GeV/$c$, indicating
only weak parton energy loss. 
For central collisions, the value for $R_{AA}$ is found to be 
$\sim$0.14 at $p_T=6$--$7$~GeV/$c$, which is 
smaller than at lower energies, despite the much less steeply falling
$p_T$ spectrum at the LHC. Above 7~GeV/$c$, $R_{AA}$ increases 
significantly.
The observed suppression of high $p_T$ particles 
provides evidence for strong parton energy loss and large
medium density at the LHC.

%

\section*{Acknowledgements}
The ALICE collaboration would like to thank all its engineers and technicians for their invaluable contributions to the construction of the experiment and the CERN accelerator teams for the outstanding performance of the LHC complex.
The ALICE collaboration acknowledges the following funding agencies for their support in building and
running the ALICE detector:
Calouste Gulbenkian Foundation from Lisbon and Swiss Fonds Kidagan, Armenia;
Conselho Nacional de Desenvolvimento Cient\'{\i}fico e Tecnol\'{o}gico (CNPq), Financiadora de Estudos e Projetos (FINEP),
Funda\c{c}\~{a}o de Amparo \`{a} Pesquisa do Estado de S\~{a}o Paulo (FAPESP);
National Natural Science Foundation of China (NSFC), the Chinese Ministry of Education (CMOE)
and the Ministry of Science and Technology of China (MSTC);
Ministry of Education and Youth of the Czech Republic;
Danish Natural Science Research Council, the Carlsberg Foundation and the Danish National Research Foundation;
The European Research Council under the European Community's Seventh Framework Programme;
Helsinki Institute of Physics and the Academy of Finland;
French CNRS-IN2P3, the `Region Pays de Loire', `Region Alsace', `Region Auvergne' and CEA, France;
German BMBF and the Helmholtz Association;
Greek Ministry of Research and Technology;
Hungarian OTKA and National Office for Research and Technology (NKTH);
Department of Atomic Energy and Department of Science and Technology of the Government of India;
Istituto Nazionale di Fisica Nucleare (INFN) of Italy;
MEXT Grant-in-Aid for Specially Promoted Research, Ja\-pan;
Joint Institute for Nuclear Research, Dubna;
 %
National Research Foundation of Korea (NRF);
CONACYT, DGAPA, M\'{e}xico, ALFA-EC and the HELEN Program (High-Energy physics Latin-American--European Network);
Stichting voor Fundamenteel Onderzoek der Materie (FOM) and the Nederlandse Organisatie voor Wetenschappelijk Onderzoek (NWO), Netherlands;
Research Council of Norway (NFR);
Polish Ministry of Science and Higher Education;
National Authority for Scientific Research - NASR (Autoritatea Na\c{t}ional\u{a} pentru Cercetare \c{S}tiin\c{t}ific\u{a} - ANCS);
Federal Agency of Science of the Ministry of Education and Science of Russian Federation, International Science and
Technology Center, Russian Academy of Sciences, Russian Federal Agency of Atomic Energy, Russian Federal Agency for Science and Innovations and CERN-INTAS;
Ministry of Education of Slovakia;
CIEMAT, EELA, Ministerio de Educaci\'{o}n y Ciencia of Spain, Xunta de Galicia (Conseller\'{\i}a de Educaci\'{o}n),
CEA\-DEN, Cubaenerg\'{\i}a, Cuba, and IAEA (International Atomic Energy Agency);
The Ministry of Science and Technology and the National Research Foundation (NRF), South Africa;
Swedish Reseach Council (VR) and Knut $\&$ Alice Wallenberg Foundation (KAW);
Ukraine Ministry of Education and Science;
United Kingdom Science and Technology Facilities Council (STFC);
The United States Department of Energy, the United States National
Science Foundation, the State of Texas, and the State of Ohio.

\newpage

\appendix
\section{The ALICE Collaboration}
\label{app:collab}
%
\begingroup
\small
\begin{flushleft}
K.~Aamodt\Irefn{0}\And
A.~Abrahantes~Quintana\Irefn{1}\And
D.~Adamov\'{a}\Irefn{2}\And
A.M.~Adare\Irefn{3}\And
M.M.~Aggarwal\Irefn{4}\And
G.~Aglieri~Rinella\Irefn{5}\And
A.G.~Agocs\Irefn{6}\And
S.~Aguilar~Salazar\Irefn{7}\And
Z.~Ahammed\Irefn{8}\And
N.~Ahmad\Irefn{9}
% ...
\renewcommand\labelenumi{\textsuperscript{\theenumi}~}
\section*{Affiliation notes}
\renewcommand\theenumi{\roman{enumi}}
\begin{Authlist}
\item \Adef{0}Also at Laboratoire de Physique Corpusculaire (LPC), Clermont Universit\'{e}, Universit\'{e} Blaise Pascal, CNRS--IN2P3, Clermont-Ferrand, France
\item \Adef{1}Now at Centro Fermi -- Centro Studi e Ricerche e Museo Storico della Fisica ``Enrico Fermi'', Rome, Italy
\item \Adef{2}Now at Physikalisches Institut, Ruprecht-Karls-Universit\"{a}t Heidelberg, Heidelberg, Germany
\item \Adef{3}Now at Frankfurt Institute for Advanced Studies, Johann Wolfgang Goethe-Universit\"{a}t Frankfurt, Frankfurt, Germany
\item \Adef{4}Now at Sezione INFN, Turin, Italy
\item \Adef{5}Now at University of Houston, Houston, Texas, United States
\item \Adef{6}Also at  Dipartimento di Fisica dell'Universit\'{a}, Udine, Italy 
\item \Adef{7}Now at SUBATECH, Ecole des Mines de Nantes, Universit\'{e} de Nantes, CNRS-IN2P3, Nantes, France
\item \Adef{8}Now at Centro de Investigaci\'{o}n y de Estudios Avanzados (CINVESTAV), Mexico City and M\'{e}rida, Mexico
\item \Adef{9}Now at Benem\'{e}rita Universidad Aut\'{o}noma de Puebla, Puebla, Mexico
\item \Adef{10}Now at Laboratoire de Physique Subatomique et de Cosmologie (LPSC), Universit\'{e} Joseph Fourier, CNRS-IN2P3, Institut Polytechnique de Grenoble, Grenoble, France
\item \Adef{11}Now at Institut Pluridisciplinaire Hubert Curien (IPHC), Universit\'{e} de Strasbourg, CNRS-IN2P3, Strasbourg, France
\item \Adef{12}Now at Sezione INFN, Padova, Italy
\item \Adef{13} Deceased 
\item \Adef{14}Also at Division of Experimental High Energy Physics, University of Lund, Lund, Sweden
\item \Adef{15}Also at  University of Technology and Austrian Academy of Sciences, Vienna, Austria 
\item \Adef{16}Also at European Organization for Nuclear Research (CERN), Geneva, Switzerland
\item \Adef{17}Now at Oak Ridge National Laboratory, Oak Ridge, Tennessee, United States
\item \Adef{18}Now at European Organization for Nuclear Research (CERN), Geneva, Switzerland
\item \Adef{19}Also at Wayne State University, Detroit, Michigan, United States
\item \Adef{20}Also at Frankfurt Institute for Advanced Studies, Johann Wolfgang Goethe-Universit\"{a}t Frankfurt, Frankfurt, Germany
\item \Adef{21}Now at Research Division and ExtreMe Matter Institute EMMI, GSI Helmholtzzentrum f\"ur Schwerionenforschung, Darmstadt, Germany
\item \Adef{22}Also at Fachhochschule K\"{o}ln, K\"{o}ln, Germany
\item \Adef{23}Also at Institute of Experimental Physics, Slovak Academy of Sciences, Ko\v{s}ice, Slovakia
\item \Adef{24}Now at Instituto de Ciencias Nucleares, Universidad Nacional Aut\'{o}noma de M\'{e}xico, Mexico City, Mexico
\item \Adef{25}Also at Laboratoire de Physique Subatomique et de Cosmologie (LPSC), Universit\'{e} Joseph Fourier, CNRS-IN2P3, Institut Polytechnique de Grenoble, Grenoble, France
\item \Adef{26}Also at  "Vin\v{c}a" Institute of Nuclear Sciences, Belgrade, Serbia 
\item \Adef{27}Also at University of Houston, Houston, Texas, United States
\item \Adef{28}Also at Department of Physics, University of Oslo, Oslo, Norway
\item \Adef{29}Also at Variable Energy Cyclotron Centre, Kolkata, India
\item \Adef{30}Now at Department of Physics, University of Oslo, Oslo, Norway
\item \Adef{31}Also at Dipartimento Interateneo di Fisica `M.~Merlin' and Sezione INFN, Bari, Italy
\item \Adef{32}Now at Nikhef, National Institute for Subatomic Physics and Institute for Subatomic Physics of Utrecht University, Utrecht, Netherlands
\item \Adef{33}Also at Hua-Zhong Normal University, Wuhan, China
\item \Adef{34}Also at Centro Fermi -- Centro Studi e Ricerche e Museo Storico della Fisica ``Enrico Fermi'', Rome, Italy
\end{Authlist}
\section*{Collaboration Institutes}
\renewcommand\theenumi{\arabic{enumi}~}
\begin{Authlist}
\item \Idef{0}Department of Physics and Technology, University of Bergen, Bergen, Norway
\item \Idef{1}Centro de Aplicaciones Tecnol\'{o}gicas y Desarrollo Nuclear (CEADEN), Havana, Cuba
\item \Idef{2}Nuclear Physics Institute, Academy of Sciences of the Czech Republic, \v{R}e\v{z} u Prahy, Czech Republic
\item \Idef{3}Yale University, New Haven, Connecticut, United States
\item \Idef{4}Physics Department, Panjab University, Chandigarh, India
\item \Idef{5}European Organization for Nuclear Research (CERN), Geneva, Switzerland
\item \Idef{6}KFKI Research Institute for Particle and Nuclear Physics, Hungarian Academy of Sciences, Budapest, Hungary
\item \Idef{7}Instituto de F\'{\i}sica, Universidad Nacional Aut\'{o}noma de M\'{e}xico, Mexico City, Mexico
\item \Idef{8}Variable Energy Cyclotron Centre, Kolkata, India
\item \Idef{9}Department of Physics Aligarh Muslim University, Aligarh, India
\item \Idef{10}Gangneung-Wonju National University, Gangneung, South Korea
\item \Idef{11}Institute for Theoretical and Experimental Physics, Moscow, Russia
\item \Idef{12}Russian Research Centre Kurchatov Institute, Moscow, Russia
\item \Idef{13}Sezione INFN, Turin, Italy
\item \Idef{14}Dipartimento di Fisica dell'Universit\`{a} and Sezione INFN, Bologna, Italy
\item \Idef{15}Bogolyubov Institute for Theoretical Physics, Kiev, Ukraine
\item \Idef{16}Frankfurt Institute for Advanced Studies, Johann Wolfgang Goethe-Universit\"{a}t Frankfurt, Frankfurt, Germany
\item \Idef{17}Dipartimento Interateneo di Fisica `M.~Merlin' and Sezione INFN, Bari, Italy
\item \Idef{18}Research Division and ExtreMe Matter Institute EMMI, GSI Helmholtzzentrum f\"ur Schwerionenforschung, Darmstadt, Germany
\item \Idef{19}V.~Fock Institute for Physics, St. Petersburg State University, St. Petersburg, Russia
\item \Idef{20}National Institute for Physics and Nuclear Engineering, Bucharest, Romania
\item \Idef{21}Kirchhoff-Institut f\"{u}r Physik, Ruprecht-Karls-Universit\"{a}t Heidelberg, Heidelberg, Germany
\item \Idef{22}Department of Physics, Ohio State University, Columbus, Ohio, United States
\item \Idef{23}Rudjer Bo\v{s}kovi\'{c} Institute, Zagreb, Croatia
\item \Idef{24}Dipartimento di Fisica dell'Universit\`{a} and Sezione INFN, Padova, Italy
\item \Idef{25}Sezione INFN, Bologna, Italy
\item \Idef{26}SUBATECH, Ecole des Mines de Nantes, Universit\'{e} de Nantes, CNRS-IN2P3, Nantes, France
\item \Idef{27}Institut f\"{u}r Kernphysik, Johann Wolfgang Goethe-Universit\"{a}t Frankfurt, Frankfurt, Germany
\item \Idef{28}Laboratoire de Physique Subatomique et de Cosmologie (LPSC), Universit\'{e} Joseph Fourier, CNRS-IN2P3, Institut Polytechnique de Grenoble, Grenoble, France
\item \Idef{29}Departamento de F\'{\i}sica de Part\'{\i}culas and IGFAE, Universidad de Santiago de Compostela, Santiago de Compostela, Spain
\item \Idef{30}Oak Ridge National Laboratory, Oak Ridge, Tennessee, United States
\item \Idef{31}Helsinki Institute of Physics (HIP) and University of Jyv\"{a}skyl\"{a}, Jyv\"{a}skyl\"{a}, Finland
\item \Idef{32}Sezione INFN, Catania, Italy
\item \Idef{33}Dipartimento di Fisica Sperimentale dell'Universit\`{a} and Sezione INFN, Turin, Italy
\item \Idef{34}Centro Fermi -- Centro Studi e Ricerche e Museo Storico della Fisica ``Enrico Fermi'', Rome, Italy
\item \Idef{35}Commissariat \`{a} l'Energie Atomique, IRFU, Saclay, France
\item \Idef{36}Laboratoire de Physique Corpusculaire (LPC), Clermont Universit\'{e}, Universit\'{e} Blaise Pascal, CNRS--IN2P3, Clermont-Ferrand, France
\item \Idef{37}Institute of Experimental Physics, Slovak Academy of Sciences, Ko\v{s}ice, Slovakia
\item \Idef{38}Dipartimento di Fisica e Astronomia dell'Universit\`{a} and Sezione INFN, Catania, Italy
\item \Idef{39}School of Physics and Astronomy, University of Birmingham, Birmingham, United Kingdom
\item \Idef{40}The Henryk Niewodniczanski Institute of Nuclear Physics, Polish Academy of Sciences, Cracow, Poland
\item \Idef{41}Institut f\"{u}r Kernphysik, Westf\"{a}lische Wilhelms-Universit\"{a}t M\"{u}nster, M\"{u}nster, Germany
\item \Idef{42}Joint Institute for Nuclear Research (JINR), Dubna, Russia
\item \Idef{43}Niels Bohr Institute, University of Copenhagen, Copenhagen, Denmark
\item \Idef{44}Institut Pluridisciplinaire Hubert Curien (IPHC), Universit\'{e} de Strasbourg, CNRS-IN2P3, Strasbourg, France
\item \Idef{45}Wayne State University, Detroit, Michigan, United States
\item \Idef{46}Petersburg Nuclear Physics Institute, Gatchina, Russia
\item \Idef{47}Physics Department, University of Jammu, Jammu, India
\item \Idef{48}Laboratori Nazionali di Frascati, INFN, Frascati, Italy
\item \Idef{49}Faculty of Nuclear Sciences and Physical Engineering, Czech Technical University in Prague, Prague, Czech Republic
\item \Idef{50}Nikhef, National Institute for Subatomic Physics, Amsterdam, Netherlands
\item \Idef{51}Centro de Investigaciones Energ\'{e}ticas Medioambientales y Tecnol\'{o}gicas (CIEMAT), Madrid, Spain
\item \Idef{52}University of Houston, Houston, Texas, United States
\item \Idef{53}Moscow Engineering Physics Institute, Moscow, Russia
\item \Idef{54}Institute for High Energy Physics, Protvino, Russia
\item \Idef{55}Faculty of Science, P.J.~\v{S}af\'{a}rik University, Ko\v{s}ice, Slovakia
\item \Idef{56}Saha Institute of Nuclear Physics, Kolkata, India
\item \Idef{57}Institut de Physique Nucl\'{e}aire d'Orsay (IPNO), Universit\'{e} Paris-Sud, CNRS-IN2P3, Orsay, France
\item \Idef{58}Department of Physics, University of Oslo, Oslo, Norway
\item \Idef{59}Dipartimento di Fisica dell'Universit\`{a} and Sezione INFN, Trieste, Italy
\item \Idef{60}Faculty of Mathematics, Physics and Informatics, Comenius University, Bratislava, Slovakia
\item \Idef{61}Russian Federal Nuclear Center (VNIIEF), Sarov, Russia
\item \Idef{62}Physikalisches Institut, Ruprecht-Karls-Universit\"{a}t Heidelberg, Heidelberg, Germany
\item \Idef{63}Physics Department, University of Cape Town, iThemba Laboratories, Cape Town, South Africa
\item \Idef{64}Hua-Zhong Normal University, Wuhan, China
\item \Idef{65}Secci\'{o}n F\'{\i}sica, Departamento de Ciencias, Pontificia Universidad Cat\'{o}lica del Per\'{u}, Lima, Peru
\item \Idef{66}Physics Department, Creighton University, Omaha, Nebraska, United States
\item \Idef{67}Universit\'{e} de Lyon, Universit\'{e} Lyon 1, CNRS/IN2P3, IPN-Lyon, Villeurbanne, France
\item \Idef{68}Universidade Estadual de Campinas (UNICAMP), Campinas, Brazil
\item \Idef{69}Nikhef, National Institute for Subatomic Physics and Institute for Subatomic Physics of Utrecht University, Utrecht, Netherlands
\item \Idef{70}Division of Experimental High Energy Physics, University of Lund, Lund, Sweden
\item \Idef{71}University of Tsukuba, Tsukuba, Japan
\item \Idef{72}Sezione INFN, Cagliari, Italy
\item \Idef{73}Centro de Investigaci\'{o}n y de Estudios Avanzados (CINVESTAV), Mexico City and M\'{e}rida, Mexico
\item \Idef{74}Benem\'{e}rita Universidad Aut\'{o}noma de Puebla, Puebla, Mexico
\item \Idef{75}Dipartimento di Scienze e Tecnologie Avanzate dell'Universit\`{a} del Piemonte Orientale and Gruppo Collegato INFN, Alessandria, Italy
\item \Idef{76}Instituto de Ciencias Nucleares, Universidad Nacional Aut\'{o}noma de M\'{e}xico, Mexico City, Mexico
\item \Idef{77}Laboratori Nazionali di Legnaro, INFN, Legnaro, Italy
\item \Idef{78}Institute of Space Sciences (ISS), Bucharest, Romania
\item \Idef{79}Institute of Physics, Bhubaneswar, India
\item \Idef{80}Universidade de S\~{a}o Paulo (USP), S\~{a}o Paulo, Brazil
\item \Idef{81}Dipartimento di Fisica `E.R.~Caianiello' dell'Universit\`{a} and Gruppo Collegato INFN, Salerno, Italy
\item \Idef{82}Sezione INFN, Bari, Italy
\item \Idef{83}Dipartimento di Fisica dell'Universit\`{a} and Sezione INFN, Cagliari, Italy
\item \Idef{84}Soltan Institute for Nuclear Studies, Warsaw, Poland
\item \Idef{85}Sezione INFN, Rome, Italy
\item \Idef{86}Faculty of Engineering, Bergen University College, Bergen, Norway
\item \Idef{87}Sezione INFN, Padova, Italy
\item \Idef{88}Institute for Nuclear Research, Academy of Sciences, Moscow, Russia
\item \Idef{89}Sezione INFN, Trieste, Italy
\item \Idef{90}Physics Department, University of Athens, Athens, Greece
\item \Idef{91}Warsaw University of Technology, Warsaw, Poland
\item \Idef{92}Universidad Aut\'{o}noma de Sinaloa, Culiac\'{a}n, Mexico
\item \Idef{93}Technical University of Split FESB, Split, Croatia
\item \Idef{94}Yerevan Physics Institute, Yerevan, Armenia
\item \Idef{95}University of Tokyo, Tokyo, Japan
\item \Idef{96}Department of Physics, Sejong University, Seoul, South Korea
\item \Idef{97}Lawrence Berkeley National Laboratory, Berkeley, California, United States
\item \Idef{98}Indian Institute of Technology, Mumbai, India
\item \Idef{99}Institut f\"{u}r Kernphysik, Technische Universit\"{a}t Darmstadt, Darmstadt, Germany
\item \Idef{100}Yonsei University, Seoul, South Korea
\item \Idef{101}Zentrum f\"{u}r Technologietransfer und Telekommunikation (ZTT), Fachhochschule Worms, Worms, Germany
\item \Idef{102}California Polytechnic State University, San Luis Obispo, California, United States
\item \Idef{103}China Institute of Atomic Energy, Beijing, China
\item \Idef{104}Institute of Physics, Academy of Sciences of the Czech Republic, Prague, Czech Republic
\item \Idef{105}University of Tennessee, Knoxville, Tennessee, United States
\item \Idef{106}Dipartimento di Fisica dell'Universit\`{a} `La Sapienza' and Sezione INFN, Rome, Italy
\item \Idef{107}Hiroshima University, Hiroshima, Japan
\item \Idef{108}Lawrence Livermore National Laboratory, Livermore, California, United States
\item \Idef{109}Budker Institute for Nuclear Physics, Novosibirsk, Russia
\item \Idef{110}Physics Department, University of Rajasthan, Jaipur, India
\item \Idef{111}Purdue University, West Lafayette, Indiana, United States
\item \Idef{112}Centre de Calcul de l'IN2P3, Villeurbanne, France 
\item \Idef{113}Pusan National University, Pusan, South Korea
\end{Authlist}
\endgroup


\begin{thebibliography}{9}
%
\bibitem{LHC} L. Evans and P. Bryant (editors), JINST {\bf 3} (2008) S08001.
\bibitem{WPBrahms} I. Arsene {\em et al.} [BRAHMS Collaboration], 
Nucl. Phys. A {\bf 757} (2005) 1.
\bibitem{WPPhobos} B.B. Back {\em et al.} [PHOBOS Collaboration], 
Nucl. Phys. A {\bf 757} (2005) 28.
\bibitem{WPStar} J. Adams {\em et al.} [STAR Collaboration], 
Nucl. Phys. A {\bf 757} (2005) 102.
\bibitem{WPPhenix} K. Adcox {\em et al.} [PHENIX Collaboration], 
Nucl. Phys. A {\bf 757} (2005) 184.
\bibitem{wang}  E.~Wang and X.~N.~Wang,
Phys.\ Rev.\ Lett.\  {\bf 89} (2002) 162301.
\bibitem{Baier} R.~Baier and D.~Schiff,
 JHEP {\bf 0609} (2006) 059.
\bibitem{wicks}
  S.~Wicks, W.~Horowitz, M.~Djordjevic and M.~Gyulassy,
  Nucl.\ Phys.\  A {\bf 784} (2007) 426.
\bibitem{vitev} I.~Vitev,
  Phys.\ Lett.\  B {\bf 639} (2006) 38.
\bibitem{eskola}
  K.~J.~Eskola, H.~Honkanen, C.~A.~Salgado and U.~A.~Wiedemann,'
  Nucl.\ Phys.\  A {\bf 747} (2005) 511.
\bibitem{glauber} M. Miller, K. Reygers, S. Sanders, and P. Steinberg,
Ann. Rev. Nucl. Part. Sci. {\bf 57} (2007) 205.
\bibitem{ALICE-det} K. Aamodt {\em et al.} [ALICE Collaboration],
JINST {\bf 3} (2008) S08002.
\bibitem{TPC-nim} J. Alme {\em et al.} [ALICE Collaboration],
Nucl. Instrum. Meth. A {\bf 622} (2010) 316.
\bibitem{paper-mult1} K. Aamodt {\em et al.} [ALICE Collaboration],
{\em accepted for publication in Phys. Rev. Lett.}, arXiv:1011.3916 [nucl-ex]
and K. Aamodt {\em et al.}, [ALICE Collaboration] {\em to be published in
Phys. Rev. Lett.}.
\bibitem{WS-parameters} H. de Vries, C.W. De Jager and C. de Vries,
Atomic Data and Nuclear Tables, {\bf 36} (1987) 495.
\bibitem{hijing} X.-N. Wang and M. Gyulassy, Phys. Rev. D {\bf 44} (1991) 3501; W.-T. Deng, X.-N. Wang, and R. Xu, (2010), 
arXiv:1008.1841 [hep-ph].
\bibitem{geant} R. Brun {\em et al.}, CERN Program Library Long Write-up,
W5013, GEANT Detector Description and Simulation Tool (1994).
\bibitem{aliroot} R. Brun {\em et al.}, [ALICE Collaboration],
Nucl. Instrum. Meth. A {\bf 502} (2003) 339.
\bibitem{paper3} K. Aamodt {\em et al.} [ALICE Collaboration],
Phys. Lett. B {\bf 693} (2010) 53.
\bibitem{7TeV} ALICE Collaboration, 
{\em to be published}.
\bibitem{hagedorn} R. Hagedorn, Riv. Nuovo Cim. {\bf 6} (1983) 1.
\bibitem{phojet} R. Engel, J. Ranft, S. Roesler, Phys. Rev. D{\bf 52} 
		(1995) 1459.
\bibitem{pythia} T. Sj\"{o}strand, S. Mrenna, P. Skands, 
		J. High Energy Phys. 0605 (2006) 026. 
\bibitem{d6t} M. Albrow et al., Tevatron-for-LHC Conference Report 
of the QCD Working Group, Fermilab-Conf-06-359, hep-ph/0610012;
T. Sj\"{o}strand and P. Skands, Eur. Phys. J. C {\bf 39} (2005) 129.
\bibitem{perugia0} P. Skands, Contribution to the 1st International Workshop on Multiple 
Partonic Interactions at the LHC, Perugia, Italy, Oct. 2008, 
Fermilab-Conf-09-113-T, arXiv:0905.3418 [hep-ph].
\bibitem{cdf} T. Aaltonen {\em et al.} [CDF Collaboration],
Phys. Rev. D {\bf 79} (2009) 112005.
\bibitem{stratmann} R. Sassot, P. Zurita, and M. Stratmann, 
Phys. Rev. D {\bf 82} (2010) 074011.
\bibitem{Abelev:2006jr} B.~I.~Abelev {\it et al.}  [STAR Collaboration],
  Phys.\ Rev.\ Lett.\  {\bf 97} (2006) 152301.
\bibitem{Adler:2003cb} S.~S.~Adler {\it et al.}  [PHENIX Collaboration],
  Phys.\ Rev.\  C {\bf 69} (2004) 034909.
\bibitem{Adler:2003au} S.~S.~Adler {\it et al.}  [PHENIX Collaboration],
  Phys.\ Rev.\  C {\bf 69} (2004) 034910.
\bibitem{Adams:2003kv} J.~Adams {\it et al.}  [STAR Collaboration],
  Phys.\ Rev.\ Lett.\  {\bf 91} (2003) 172302.
\bibitem{Fries:2003fr} R.~J.~Fries and B.~Muller,
  Eur.\ Phys.\ J.\  C {\bf 34} (2004) S279.

\end{thebibliography}
\end{document}